\documentclass{PoS}
\usepackage[outercaption]{sidecap}

\title{Measurement of forward direct photon
production in p--A at the LHC with ALICE
 --
\\A probe for nuclear PDFs and saturation}

\ShortTitle{Forward Direct Photons in p--A with ALICE}

\author{\speaker{Thomas Peitzmann}\thanks{for the ALICE FoCal Collaboration}\\
        Utrecht University/Nikhef, The Netherlands\\
        E-mail: \email{t.peitzmann@uu.nl}}

\abstract{Probes for the small-$x$ parton densities and predicted effects of gluon saturation are discussed. At very low $x$ and intermediate $Q$, only results on hadronic observables at the LHC are available, which do not provide unambiguous information. It is shown that the measurement of direct photons at forward rapidity at the LHC is particularly promising to provide a unique signal. We further discuss the possibilities to perform such measurements with a detector upgrade in the ALICE experiment and present the R\&D activities ongoing.}

\FullConference{XXIV International Workshop on Deep-Inelastic Scattering and Related Subjects\\
		11-15 April, 2016\\
		DESY Hamburg, Germany}

\begin{document}

\section{Small-$x$ parton distributions and gluon saturation -- a status}
While the parton densities at small momentum fraction $x$ in hadrons and nuclei are crucial for the quantitative description of hadron collisions at high energy, 
they are only very badly known.  At moderate $Q^2$, there are essentially no direct experimental constraints of the PDFs for protons at $x < 10^{-4}$ and for nuclei at $x < 10^{-2}$. Linear QCD evolution (DGLAP) is used to extrapolate to the inaccessible kinematic regions. However, the increase of parton densities towards low $x$ cannot continue indefinitely. Non-linear effects are expected to ``tame'' the gluon density and lead to \textit{gluon saturation} -- these effects are enhanced in nuclei compared to protons because of the higher parton density. Models of gluon saturation, like the \textit{colour glass condensate}, provide alternative quantitative descriptions of the initial state of hadrons and nuclei at very high energy \cite{CGC}. 

Gluon saturation is consistent with many phenomena observed in high-energy scattering, however, no unambiguous proof has been obtained so far. For many of the observables used, alternative explanations are possible. DIS is currently limited in its kinematic reach.
Measurements at hadron colliders are potentially very interesting, but so far have used hadronic observables, which do bring additional uncertainties. For example, measurements in d--Au collisions at RHIC have shown a suppression of hadron yields and dihadron correlations at forward rapidity, 
but these data are limited to very small transverse momenta, which makes the theoretical calculations using pQCD unreliable.

Measurements at the LHC offer new opportunities to study low-$x$ partons due to the high $\sqrt{s}$, as at leading order the final state parton $p_T$ and $y$ are approximately related to the initial  kinematics via $x_2 \approx 2 p_T e^{-y} / \sqrt{s} $.
Forward hadron production has been studied in p--Pb collisions at 5.02~TeV, because saturation effects should be enhanced in nuclei, while the complications from medium effects in Pb--Pb should be largely absent. A suppression of forward production has been observed for several hadron species, which may be described by a combination of cold nuclear matter effects. I will briefly discuss the two most interesting measurements in this context here.

\begin{figure}[htb]
\begin{center}
\includegraphics[width=0.45\textwidth]{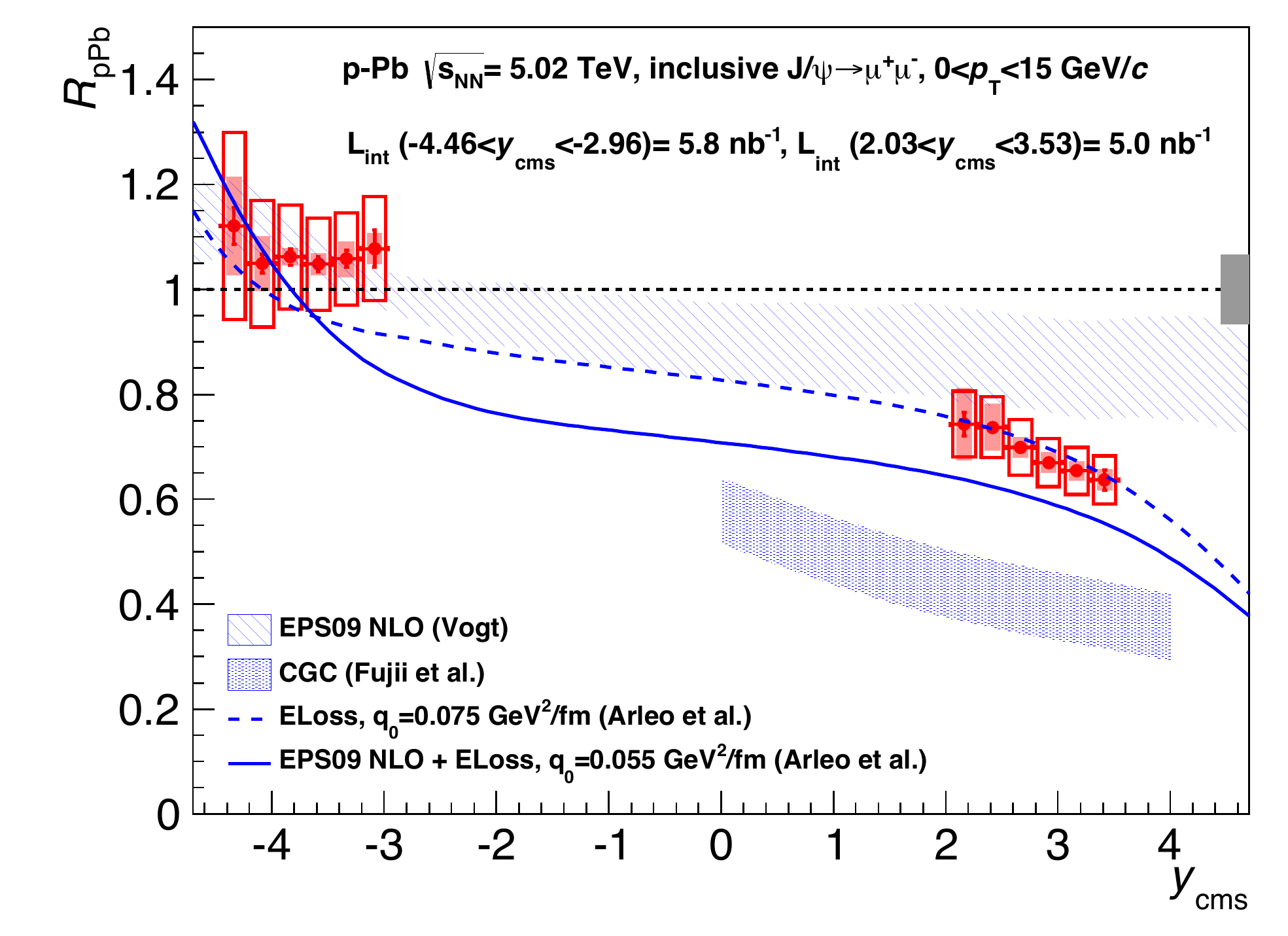}\hfill%
\includegraphics[width=0.49\textwidth]{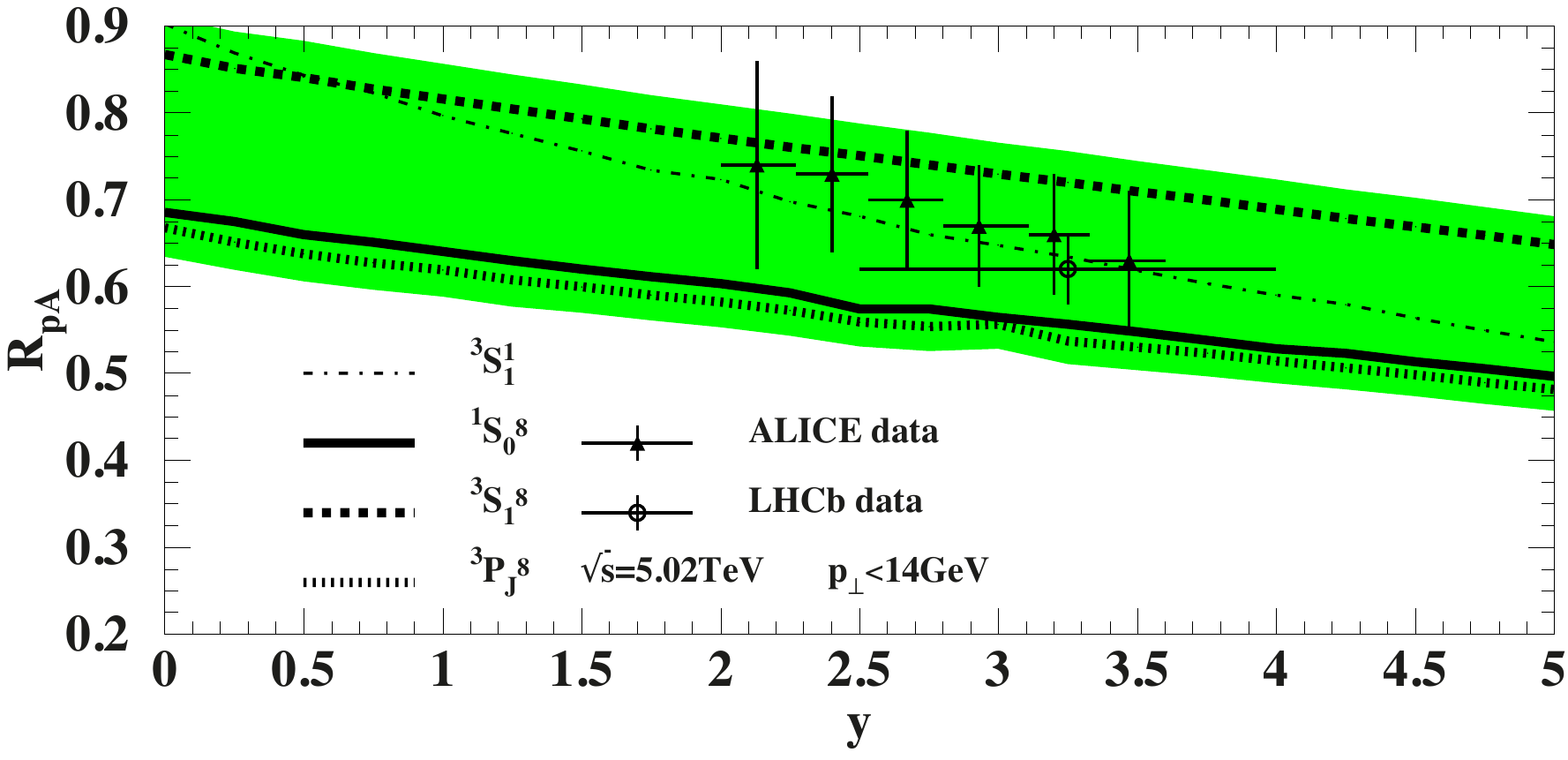}
\caption{\protect\label{fig:jpsialice}  Nuclear modification factor $R_{pPb}$ for $J/\psi$ as a function of $y$. Left: The experimental data from \cite{alice-jpsi} are shown as red symbols and are compared to several theoretical predictions.  
Right: Experimental data compared to CGC model calculations \cite{venugo-jpsi}.} 
\end{center}
\end{figure}

Fig.~\ref{fig:jpsialice} displays the nuclear modification factor:
\begin{displaymath}
R_{pPb} = \frac{ \left. d^2N/dp_T dy \right|_{pPb}}{\left\langle N_{coll} \right\rangle \left. d^2N/dp_T dy  \right|_{pp}}
\end{displaymath}
for $J/\psi$ production as a function of $y$ as measured in ALICE \cite{alice-jpsi}. At backward rapidity ($y < 0$) no significant nuclear modification is observed -- if anything, there is a slight enhancement of the yield. At forward rapidity ($y > 0$) the production of $J/\psi$ is suppressed, i.e. we find $R_{pPb} < 1$. The data are compared with theoretical calculations: An NLO pQCD calculation using realistic nuclear PDFs (EPS09 \cite{Eskola:2009uj}) including shadowing predicts too little suppression. A model calculation of energy loss in cold nuclear matter by itself is able to describe the data well; if these calculations also include shadowing, a slightly too strong suppression is obtained. Finally there is also a CGC model calculation \cite{fujii-jpsi}, which underpredicts $R_{pPb}$. However, this was a fairly simple CGC calculation, both in terms of the nuclear geometry that was used and the hadronisation model for $J/\psi$. A more advanced CGC calculation of $J/\psi$ production \cite{venugo-jpsi} is shown in the right panel of Fig.~\ref{fig:jpsialice}. 
This new calculation is fully con\-sistent with the experimental data, both in the rapidity dependence of the suppression as well as the $p_T$ dependence (not shown here). The green bands show the sizeable theoretical uncertainty, which make it impossible to obtain a conclusive statement on gluon saturation from these existing experimental results. In fact, also  the comparison of the suppression of $\psi(2S)$ to that of the $J/\psi$ \cite{alice-jpsi} hints at possible final state effects in these observables.


\begin{SCfigure}
\includegraphics[width=0.45\textwidth]{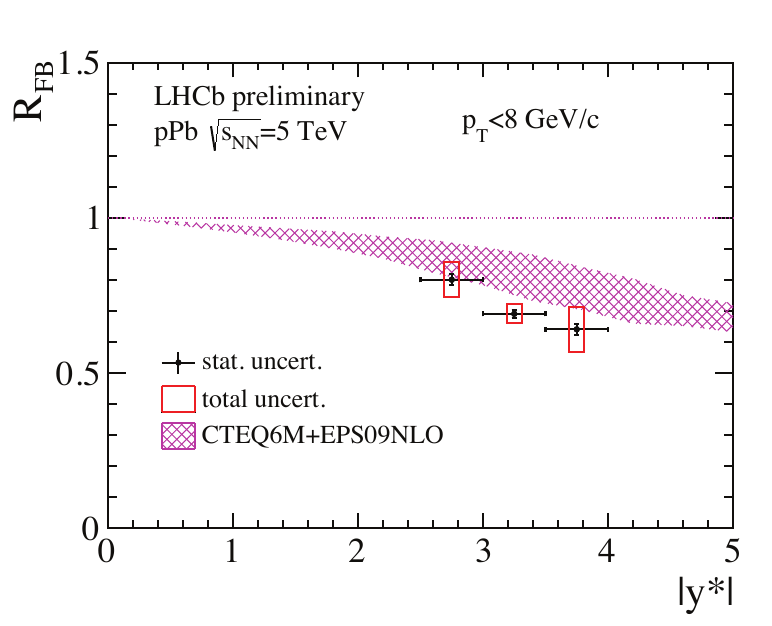}
\caption{\protect\label{fig:dlhcb} 
Forward-backward production ratio $R_{FB}$ for D mesons as a function of $y$ integrated over $p_T < 8$~GeV/$c$ for p--Pb collisions at 5~TeV as measured by LHCb (from \cite{LHCb-CONF-2016-003}). Also shown is a prediction of NLO pQCD using EPS09 nuclear PDFs.}
\end{SCfigure}

As another example LHCb has presented preliminary results of forward D-meson production in p--Pb  collisions  \cite{LHCb-CONF-2016-003}. Fig.~\ref{fig:dlhcb} shows the forward-backward production ratio $R_{FB}$ as a function of rapidity $y$. Forward D production is apparently suppressed compared to backward production, and the relative suppression grows for larger $y$. An NLO pQCD calculation including shadowing also predicts a suppression, albeit not quite as strong as seen in the data. This could be seen as an indication of additional suppression as expected from saturation. Unfortunately, the uncertainties are relatively large thus not allowing any strong conclusions. Possibly, an effect would be strong enough for a clear observation only for still larger rapidities, which are not available from LHCb measurements.

In summary, existing measurements of hadronic observables do not provide a conclusive answer on the question of gluon saturation. Likely measurements at still higher rapidity could provide clearer signals, however, a large additional uncertainty relates to possible effects of final state interactions, with possibly a yield suppression from energy loss, or even a yield enhancement, which one might expect from the first hints of collective motion that have been observed in p--Pb. In this context the only possibility to reduce the corresponding systematic uncertainties is to use probes that do not suffer strong final state interaction, namely electromagnetic probes.

\section{Forward Direct Photons and the FoCal Upgrade in ALICE}
\begin{figure}[bth]
\begin{center}
\includegraphics[width=0.4\textwidth]{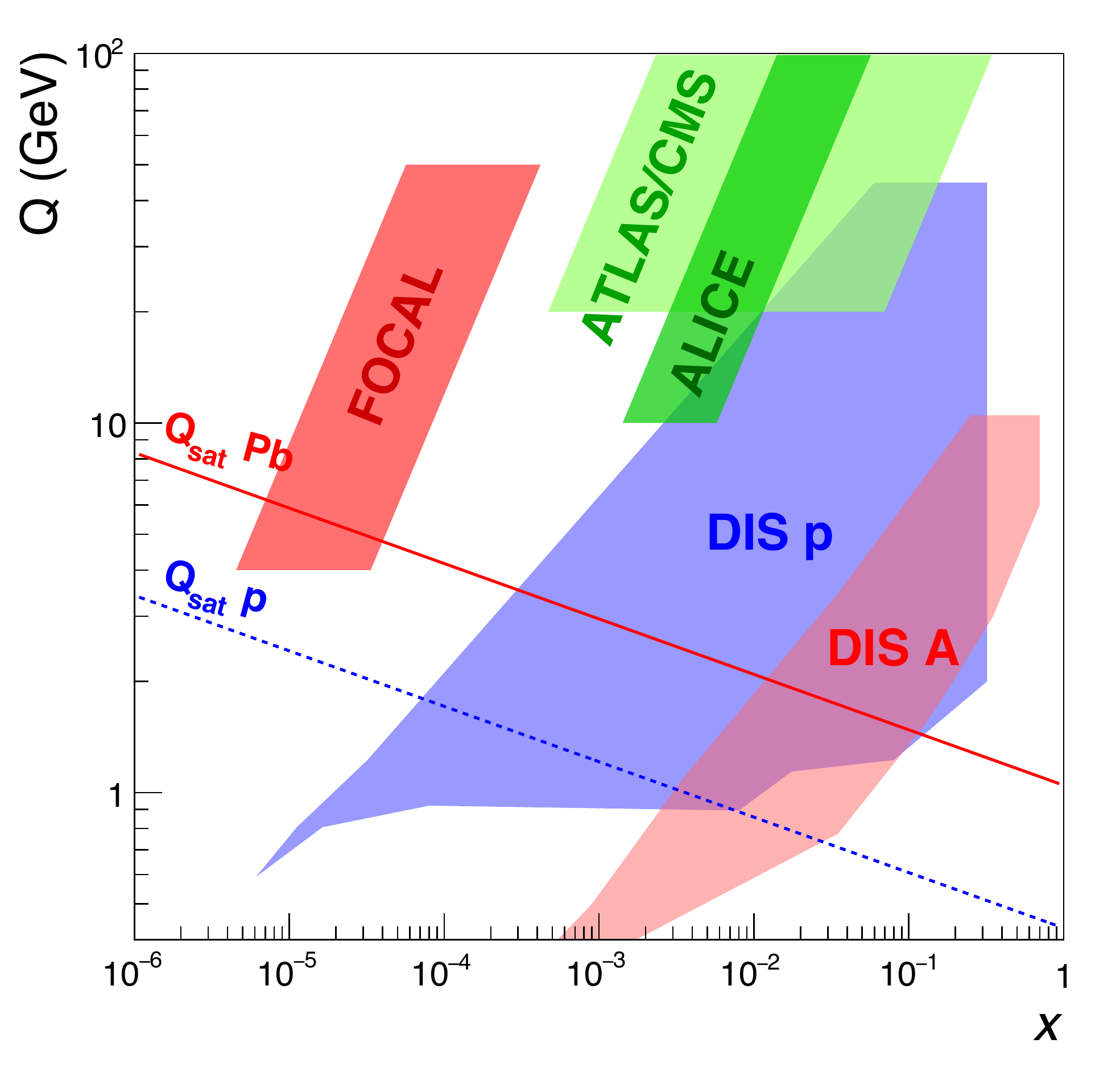}
 \includegraphics[width=0.5\textwidth]{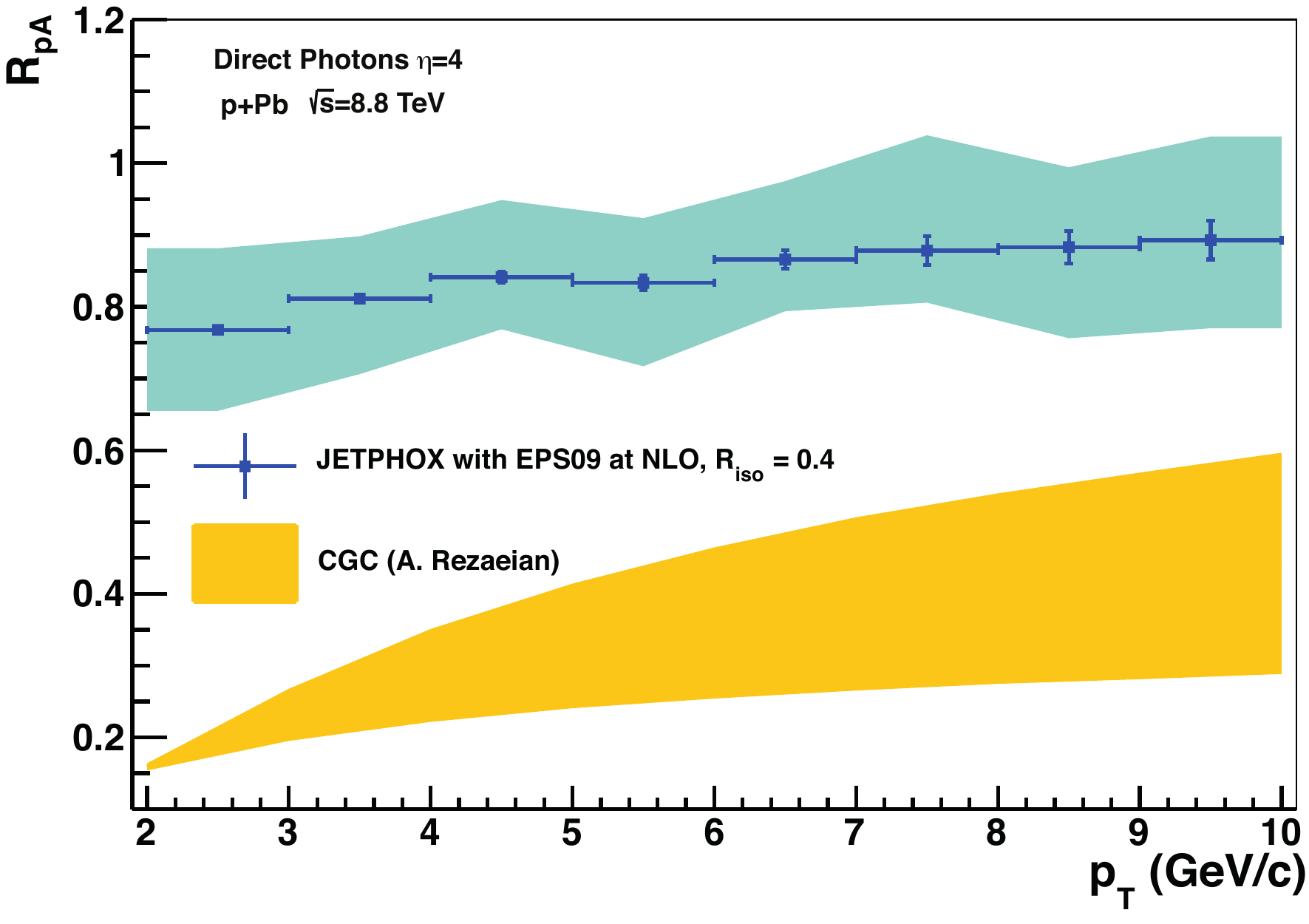}
\caption{Left: Approximate sensitivity regions in $x$ and $Q$ probed by DIS measurements and direct photon measurements at the LHC. The estimated saturation scale is also indicated. Right: Nuclear modification factor $R_{pA}$ as a function of $p_T$ for forward direct photon production. Shown are results of CGC calculations from Ref.~\cite{Rezaeian:2012} (orange) and from NLO pQCD calculations with JETPHOX (blue). The shaded bands show the systematic error estimates. \protect\label{fig:rpa}}
\end{center}
\end{figure}

At leading order, photons are directly sensitive to the gluon distributions via the quark-gluon-Compton process, and the relative contribution of higher order diagrams can be suppressed by appropriate isolation cuts. This is in principle also true for Drell-Yan pair production at next-to-leading order via the virtual Compton process, however, the much smaller cross section for Drell-Yan will not allow conclusive measurements for the luminosity expected in p--Pb collisions at the LHC. This leaves only real direct photons as a promising probe. In addition, compared to hadrons, which suffer strongly from effects of fragmentation, photons, in particular when isolated, have a much clearer sensitivity to small $x$. The left panel of Fig.~\ref{fig:rpa} shows the expected sensitivity regions in $x$ and $Q$ probed by DIS measurements and direct photon measurements at the LHC. Measurements of direct photons for $y > 4$, as possible with the proposed FoCal detector in ALICE as discussed below, should strongly constrain the gluon density in protons and nuclei down to $x = 10^{-5}$, advancing our knowledge of the proton and nuclear PDFs significantly. 

Moreover, such a measurement will have a clear sensitivity to effects of gluon saturation. For photon production, state-of-the-art calculations both of NLO pQCD and of a saturation model are available. The nuclear modification factor $R_{pPb}$ for $y = 4$ according to pQCD has been calculated with JETPHOX using EPS09 nuclear PDFs. The result for $R_{pPb}$ as a function of $p_T$ is shown in the right panel of Fig.~\ref{fig:rpa} as the blue symbols. The production is slightly suppressed to $\approx 0.8$ at low $p_T$ in p--Pb collisions due to the shadowing as incorporated in the nuclear PDFs -- the ratio slowly rises towards higher $p_T$. This is compared to the CGC calculation from \cite{Rezaeian:2012} shown as an orange band in the figure -- the width of the band reflects the systematic uncertainties of the calculations. These calculations predict a very strong suppression to $< 0.2$ for the lowest $p_T$, and even at $p_T = 5 \, \mathrm{GeV}/c$ there is still a suppression to $R_{pPb} < 0.4$.

Such measurements will not be possible with existing detectors. 
Direct photons will have to be identified at $3.5 < y < 5$ for $p_T \approx 4-20 \, \mathrm{GeV}/c$, which corresponds to photon energies of several 100 GeV, where one uses electromagnetic calorimeters. For such high energies, the decay photons from neutral pions provide a formidable background, as the typical opening angles are smaller than $10^{-3}$ radians. Even at large distances, the two resulting showers would have a separation of only a few millimeters. To reject the decay photons, a detector will have to be capable to resolve such close-by showers, which requires a detector of very small Moli\`ere radius and extremely fine granularity, beyond state-of-the-art technology.

\begin{figure}[bth]
\begin{center}
  \includegraphics[width=0.4\textwidth]{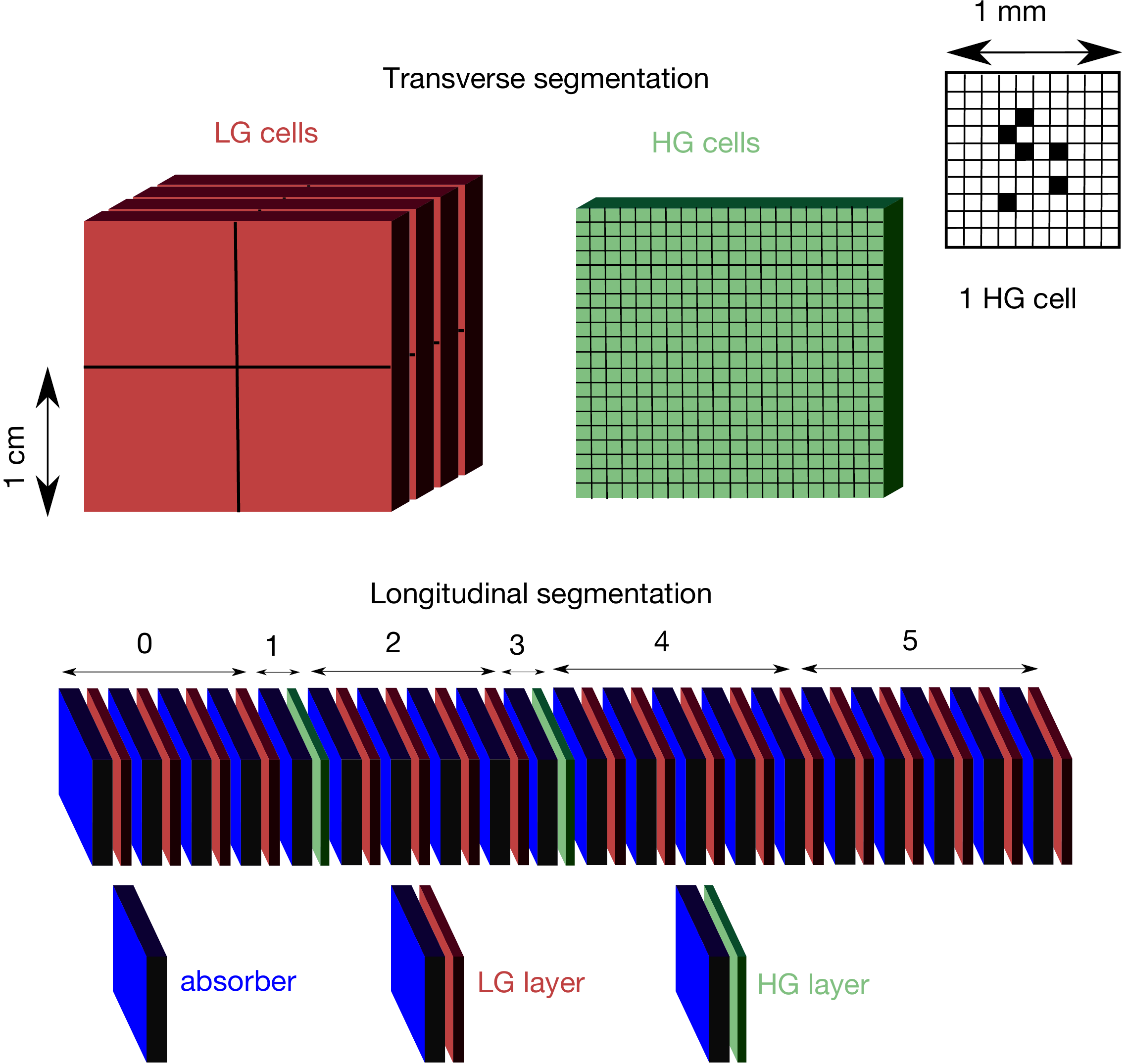}\hfill%
  \includegraphics[width=0.4\textwidth]{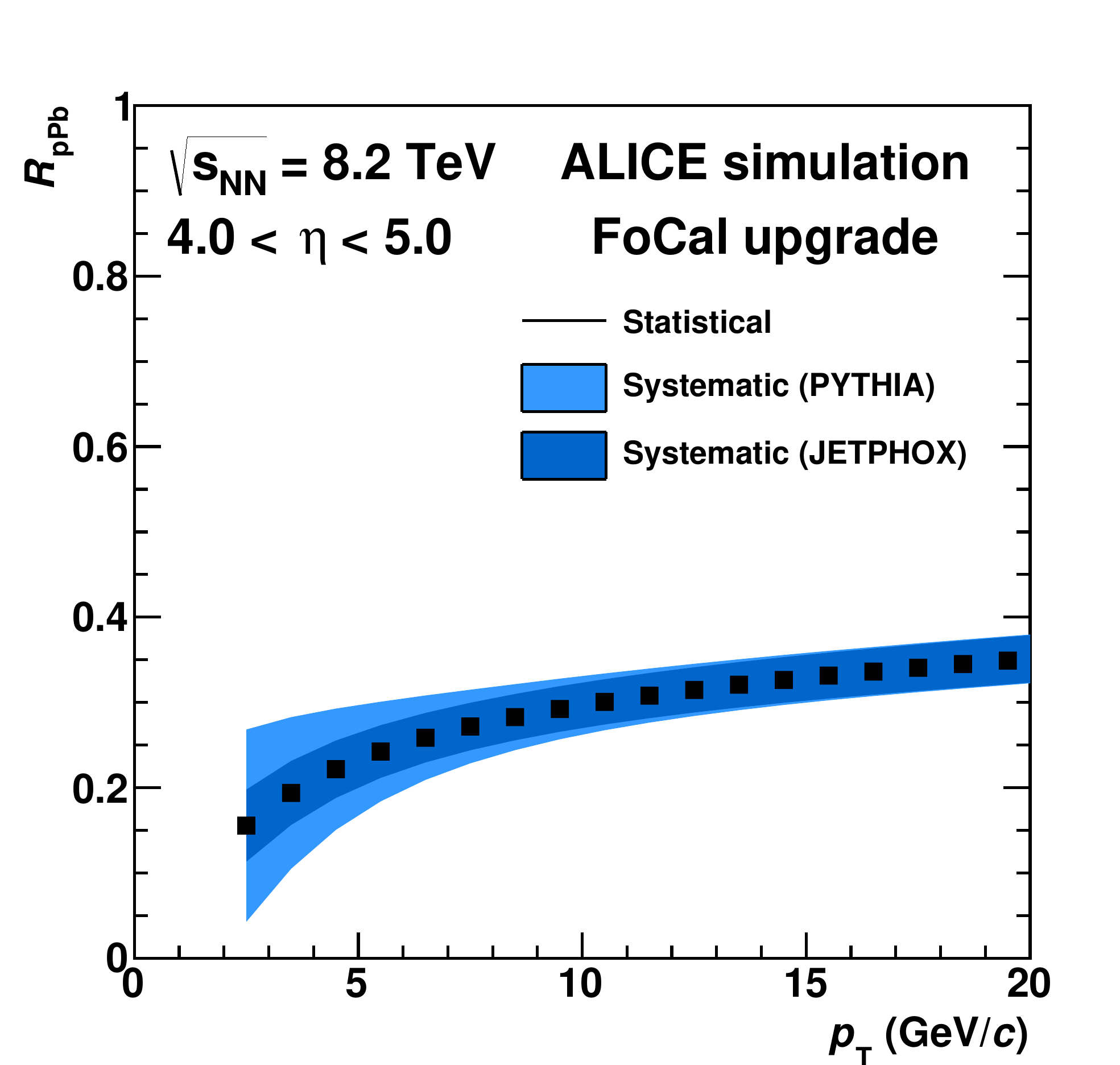}
  \caption{Left: Schematic view of the internal structure of the FoCal detector with W-absorbers and Si-sensor layers of low (LG) and high (HG) granularity. Right: Expected relative uncertainty on a direct photon $R_\mathrm{pPb}$ measurement for ${\cal{L}}_{int} = 50 \, \mathrm{nb}^{-1}$.  The bands indicate the systematic uncertainty, mostly due to uncertainties on the efficiency and energy scale, as well as the decay photon background determination.
   \label{fig-design}}
\end{center}
\end{figure}

Within the ALICE collaboration, it has been proposed to implement a forward calorimeter (FoCal) as an upgrade in the LHC long shutdown 3. The proposed detector covers the range $3.5 < \eta < 5$, which probes the gluon distributions at $x \approx 10^{-5}$ and $Q \approx p_T > 4 \, \mathrm{GeV}$. The detector would consist of a Si-W sandwich structure with a number of layers of extremely high granularity (HG), which would provide the two-shower separation capability, and layers of somewhat lower granularity (LG) more suited for the total energy measurement. A sketch of the internal structure as currently implemented in simulations is shown in the left panel of Fig.~\ref{fig-design}. Detailed performance simulations using such a detector design have demonstrated that the forward direct photon measurements should be possible, as is illustrated by the right panel of Fig.~\ref{fig-design}, which shows the expected uncertainty on the measurement of the nuclear modification factor of direct photons assuming an integrated luminosity of ${\cal{L}}_{int} = 50 \, \mathrm{nb}^{-1}$ for p--Pb.  The simulations also demonstrate that the HG layers are instrumental in obtaining this performance. Only a granularity of the order of  $\approx 1$~mm$^2$ or better allows the two-shower separation necessary for efficient rejection of high-energy pions. 

\begin{SCfigure}
\includegraphics[width=0.5\textwidth]{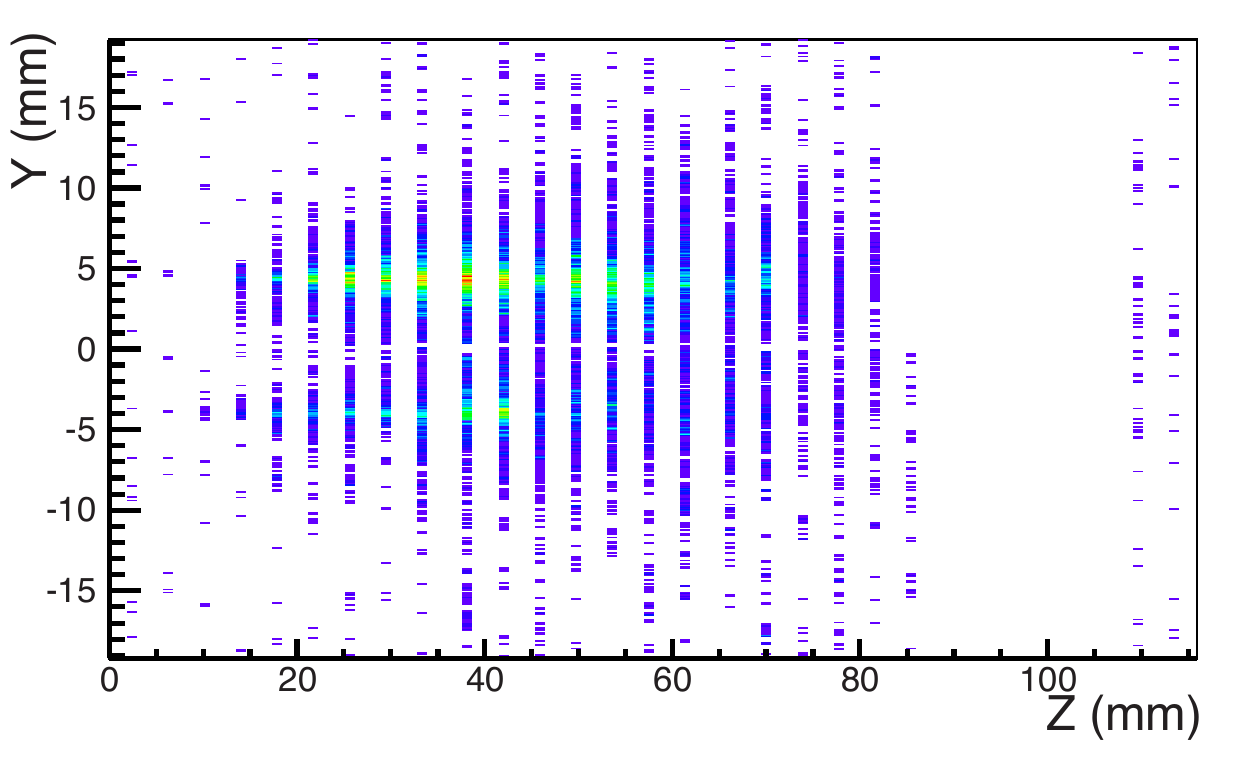}
\caption{Side view of the density distribution of fired pixels for the simultaneous measurement in the CMOS calorimeter prototype of two showers from a mixed test beam at the CERN SPS, illustrating the two shower separation capabilities.
   \label{fig-twoshower}}
\end{SCfigure}

Extensive R\&D is being performed on both the LG and HG technologies. While the LG layers will likely use Si-pad sensors, the most promising candidate for the HG layers are CMOS sensors, which allow a high pixel density with little additional material, keeping the effective Moli\`ere radius small. However, very little is known at present about their behaviour in a calorimeter, where the typical pixel occupancies are much higher than in tracking applications. A prototype of a high-granularity electromagnetic calorimeter using MIMOSA23 sensors \cite{mimosa} with a total number of 39 million pixels of $30 \times 30 \, \mu\mathrm{m}^2$ in a $4 \times 4 \times 11\,\mathrm{cm}^3$ volume and a Moli\`ere radius of $\approx 11$~mm has been built and tested successfully. The results clearly demonstrate the extraordinary capabilities of two-shower separation. Fig.~\ref{fig-twoshower} shows an event display of the measurement of two particles in the prototype, where one can clearly separate the showers -- this should be possible down to distances of a few mm. The sensor used in the prototype thus fulfils many of the crucial requirements for the FoCal upgrade and can be considered a successful proof of principle, it is however too slow to be used in a real experiment. The further R\&D will make use of synergy with the sensor development for the ALICE ITS upgrade \cite{alice-its} to obtain suitable sensors for the final detector.

In summary, the best opportunity in the near future to significantly constrain the low-$x$ structure of protons and nuclei and to shed light on the open question of gluon saturation is the measurement of direct photons at forward rapidity in p--A and pp at the LHC. While no existing detector would be able to perform this measurement, it should be possible with a proposed upgrade to the ALICE experiment with a Forward Calorimeter (FoCal). To allow this measurement, an extremely high granularity calorimeter beyond the state of the art has to be developed. Corresponding R\&D is ongoing, and the proof-of-principle measurements have been successfully performed.


\end{document}